\documentclass[twocolumn,showpacs,showkeys,a4paper,superscriptaddress,nofootinbib]{revtex4}
\usepackage{graphicx,multirow}
\usepackage[usenames,dvipsnames]{color}
\usepackage{amsfonts}
\usepackage{mathrsfs}
\usepackage[innercaption]{sidecap}
\usepackage{overpic}
\usepackage{hyperref}

\newcommand{\diff}{{\rm d}}
\newcommand{\scri}{\mathscr{I}}

\newcommand{\eplanck}{E_{\rm B}}
\newcommand{\elor}{E_{\rm L}}
\begin{document}
\setcounter{topnumber}{1}
\title{Quantum Non-Gravity and Stellar Collapse}
\author{C. Barcel\'{o}}
\affiliation{Instituto de Astrof\'{\i}sica de Andaluc\'{\i}a (IAA-CSIC),
Glorieta de la Astronom\'{\i}a, 18008 Granada, Spain}
\author{L. J. Garay}
\affiliation{Departamento de F\'{\i}sica Te\'{o}rica II, Universidad Complutense
de Madrid, 28040 Madrid, Spain}
\affiliation{Instituto de Estructura de la Materia (IEM-CSIC), Serrano 121,
28006 Madrid, Spain}
\affiliation{King's College London, Department of Physics, Strand, London WC2R
2LS, UK}
\author{G. Jannes}
\affiliation{Instituto de Astrof\'{\i}sica de Andaluc\'{\i}a (IAA-CSIC),
Glorieta de la Astronom\'{\i}a, 18008 Granada, Spain}
\affiliation{Instituto de Estructura de la Materia (IEM-CSIC), Serrano 121,
28006 Madrid, Spain}

\date{10 May 2011}

\begin{abstract}
Observational indications combined with analyses of analogue and
emergent gravity in condensed matter systems support the possibility
that there might be two distinct energy scales related to quantum
gravity: the scale that sets the onset of quantum gravitational effects
$\eplanck$ (related to the Planck scale) and the much higher scale
$\elor$ signalling the breaking of Lorentz symmetry. We suggest a
natural interpretation for these two scales: $\elor$ is the energy scale
below which a special relativistic spacetime emerges, $\eplanck$ is the
scale below which this spacetime geometry becomes curved. This implies
that the first `quantum' gravitational effect around $\eplanck$ could
simply be that gravity is progressively switched off, leaving an effective
Minkowski quantum field theory up to much higher energies of the order
of $\elor$. This scenario may have important consequences for
gravitational collapse, inasmuch as it opens up new possibilities for the
final state of stellar collapse other than an evaporating black hole.
\end{abstract}
\pacs{04.60.-m, 04.70.Dy, 97.60.Lf}
\keywords{Quantum Gravity; Black hole singularity; Emergent gravity; Fermi liquids}
\maketitle

\section{Introduction}

Indications from recent cosmic ray and other
high-energy observations show that there exist stringent bounds on the
most commonly expected types of Lorentz violation at the Planck
level~\cite{Jacobson:2005bg,Maccione:2009ju}. Two possible
interpretations stand out. Either Lorentz invariance is a truly
fundamental symmetry of our universe, valid at all energies. Or Lorentz
invariance is nevertheless violated, but at an energy $\elor$ much higher
than the Planck scale $\eplanck$.

It is thought-provoking to realize that this second option is precisely a
crucial condition for an emergent theory of gravity based on a
fermionic system (similar in some aspects to certain condensed matter systems)  to
work~\cite{Volovik:2006ft}. We will briefly describe how the Einstein equations
can be recovered within such an emergent gravity framework, and specifically for
Fermi-liquid-like systems. Then we will explore a rather surprising consequence
of this proposal: The physics at energies above the Planck scale (but still
below the much higher Lorentz violation energy scale $\elor$) could be described
essentially as standard quantum field theory in a Minkowski spacetime. This
opens up the possibility of new final stages for the gravitational collapse of
ultra-heavy objects which avoid the formation of a general relativistic
singularity. In particular, we suggest that an object similar to an isothermal
sphere could after all be recovered as a natural outcome of such a collapse.

\section{Emergent gravity, a tale of two scales}

In a Fermi liquid like the
A-phase of helium three ($^3$He-A), there exist two important energy scales
relevant for our discussion~\cite{Volovik:2003fe,Volovik:2006ft,Volovik:2007vs}.
One is the energy scale $\eplanck$ below which bosonisation in the system starts
to develop. This scale marks the onset of the superfluid behaviour of $^3$He-A.
At energies below $\eplanck$, the different bosons effectively appearing in the
system   (e.g.\ through Cooper-pairing) condense and start to exhibit collective
behaviours. The other energy scale $\elor$ is the Lorentz scale below which the
quasiparticles of the system start to behave relativistically (as Weyl spinors).
This occurs in $^3$He-A because then the  momentum space  topology of the vacuum
has Fermi points. It is in the immediate surroundings of these Fermi points that
such a relativistic behaviour appears.\footnote{Fermions obey a relativistic
Dirac equation near all types of Fermi surfaces, but only in the directions
perpendicular to the Fermi surface, as a consequence of
the Atiyah-Bott-Shapiro construction \cite{atiyah,Horava:2005jt}. In order to reproduce true Lorentz
invariance in an emergent gravity setting, the relevant topological object must
therefore be zero-dimensional, i.e.: a Fermi point.}

At energies below both characteristic scales, one can describe  the
system as a set of Weyl spinors coupled to emergent  background
electromagnetic and gravitational fields. For a particular Fermi point,
these  effective electromagnetic and gravitational fields encode,
respectively, the position of the point and its  effective  ``light-cone''
structure. Both electromagnetic and gravitational fields are built from
condensed bosonic degrees of freedom. Apart from any predetermined
dynamics, these bosonic fields will acquire additional dynamical
properties through Sakharov's induction
mechanism~\cite{Sakharov:1967pk,Visser:2002ew}. Integrating out the
effect of quantum fluctuations in the fermionic fields {\it \`{a} la} Sakharov,
one obtains a one-loop effective action for the geometric field, to be
added to any pre-existing tree level contribution. Since $\eplanck$
marks the energy scale above which the geometrical picture based on the
bosonic condensate disappears, the integration cut-off is precisely this
$\eplanck$.

Now, in order for the geometrical degrees of freedom to follow an
Einsteinian dynamics, two conditions are required:
\begin{enumerate}

\item
{\it Special relativity dominance or $\elor \gg \eplanck$}: For the
induction mechanism to lead to an Einstein-Hilbert term $\sqrt{-g}R$
in the effective Lagrangian, the fluctuating fermionic field must
``feel'' the geometry (i.e., it must fulfil a locally Lorentz-invariant
equation) at all scales up to the cut-off. The term $\sqrt{-g}R$
appears multiplied by a constant proportional to $\eplanck^2$, which
originates from an integral of the type $\int k\diff
k$~\cite{Sakharov:1967pk}. The correct value for Newton's constant
$G$ in the Einstein-Hilbert action is then recovered precisely if one
identifies the  cut-off energy $\eplanck$ with the Planck energy. Now,
this $(\int k\diff k)$-dependence of the gravitational coupling
constant tells us that the quantum fluctuations which are most relevant
in producing the Einstein-Hilbert term are those with energies close to
the cut-off, that is, around the Planck scale. Therefore, to ensure the
induction of an Einstein-Hilbert term, these fermionic fluctuations
with energies close to the Planck scale must be perfectly Lorentzian to
a high degree. This can only be ensured if $\elor \gg \eplanck$.

\item
{\it Sakharov one loop dominance}: One also needs that the induced
dynamical term dominates over any pre-existing tree level contribution.

\end{enumerate}

Unfortunately, such special relativity dominance is not realised in $^3$He-A,
where the opposite happens: \mbox{$\eplanck \gg \elor$}, nor in any other known
condensed matter system~\cite{Barcelo:2005fc}. Therefore the effective dynamics
of the gravitational degrees of freedom emerging in real laboratory condensed
matter systems are not relativistic but of fluid-mechanical type. This crucial observation could well be related to the following. Perhaps the fact that $\elor\ll\eplanck$ in condensed matter models is related to the {\em background dependence} of these models, contrarily to what happens in certain theories
of quantum gravity such as Loop Quantum Gravity. 

But what about the gravitational
degrees of freedom of our universe? What if gravity were really emergent along
the described scenario and the previous two conditions were fulfilled? In particular, what if $\elor\gg\eplanck$, unlike what happens in any known
condensed matter system? Indeed, if gravity is to be an emergent phenomenon, its
underlying structure will not have the exact same properties as ordinary matter. The
microscopic system underlying general relativity cannot simply be a condensed matter system,
but a condensed-matter-{\em like} system, whose specific characteristics we do not
really know.

\section{The realm of quantum non-gravity }

The following conceptual image could
then follow from these considerations.  Whatever the microscopic details of the
ultra-high-energy fermionic theory of our universe (and whatever
symmetry, Galilean or otherwise, this theory may possess), as the energy
decreases below $\elor$, the degrees of freedom start to behave as fermionic spinwaves in a
Lorentz-invariant background geometry: A special relativistic spacetime emerges.
Then, as the energy further decreases below $\eplanck$, some of the fermions
couple into effective bosons which condense to provide the dynamical degrees of
freedom of the geometry: Spacetime becomes curved.
Note that since the dynamics of the spacetime is induced, the Weinberg-Witten theorem~\cite{Weinberg:1980kq} does not apply. As far as the collapse scenario that we will describe in the next section is concerned, the energy scale $\elor$ could even be infinite, so that no Lorentz invariance violations would take place. In general terms, $\elor$ plays little role in this scenario.
To emphasise the crucial point, let us invert the reasoning and increase the energy, 
starting from our low-energy world. 
Then the first `quantum' gravitational effect taking place at the Planck scale
$\eplanck$ would simply be that gravity is progressively switched
off and the curved geometry
becomes flat. One would be left with the effective paradigm of standard Effective Quantum Field
Theory in Minkowski Spacetime with an energy cutoff at $\elor$ (for related ideas, see \cite{Boughn:2008jx}). Of course, beyond
$\elor$ there would still remain some yet-to-be-discovered full-fledged theory
of `quantum gravity'.

Even the energy density in a neutron star (where the distance between
the neutrons is of the order of their Compton wavelength $10^{-15}$m) is
roughly eighty orders of magnitude below the Planck energy density.
General relativity is therefore recovered even in the extremely dense
scenarios occurring in neutron stars.  Next, we will take a speculative look at what
could happen in even denser situations.

\section{Nonsingular gravitational collapse}

\subsection{Single-bounce model}

Consider the simple case of a spherical shell of matter collapsing from an infinite radius under its own gravitational pull. We take it to be sufficiently massive to overcome the pressure
of the Pauli exclusion principle as it collapses.
As the collapse advances towards the formation of a
general-relativistic singularity, the energy and momentum of the
constituent particles are constantly increasing, taking advantage 
of the gravitational potential well.
Eventually, the shell radius approaches a critical value where the energy of the infalling particles
becomes of the order of the Planck scale
$\eplanck$. Then, within the scenario described above, gravity would be progressively
switched off and the particles just perceive the Minkowski structure
which persists up to much higher energies. This means that the total energy of the infalling particles
becomes a conserved quantity and does not further increase once beyond
$\eplanck$.

The fate of the shell is from now on governed by standard relativistic
quantum field theory. All the particles in the shell can be seen as the past
external fermionic lines of a quantum scattering process. The final result
of the collision will be another collection of particles with the same total
energy and momentum. For simplicity, let us assume that the final
collection of particles maintains the spherical shell  shape and that
dissipation is negligible. Then, after the collision, an equivalent shell could be
recovered with the same critical radius but now expanding in time. Once
the shell has expanded beyond the critical radius (or critical density), the
general relativistic notions of spacetime curvature again apply.

\begin{figure}%
\vbox{ \hfil
\includegraphics[width=.23\textwidth]{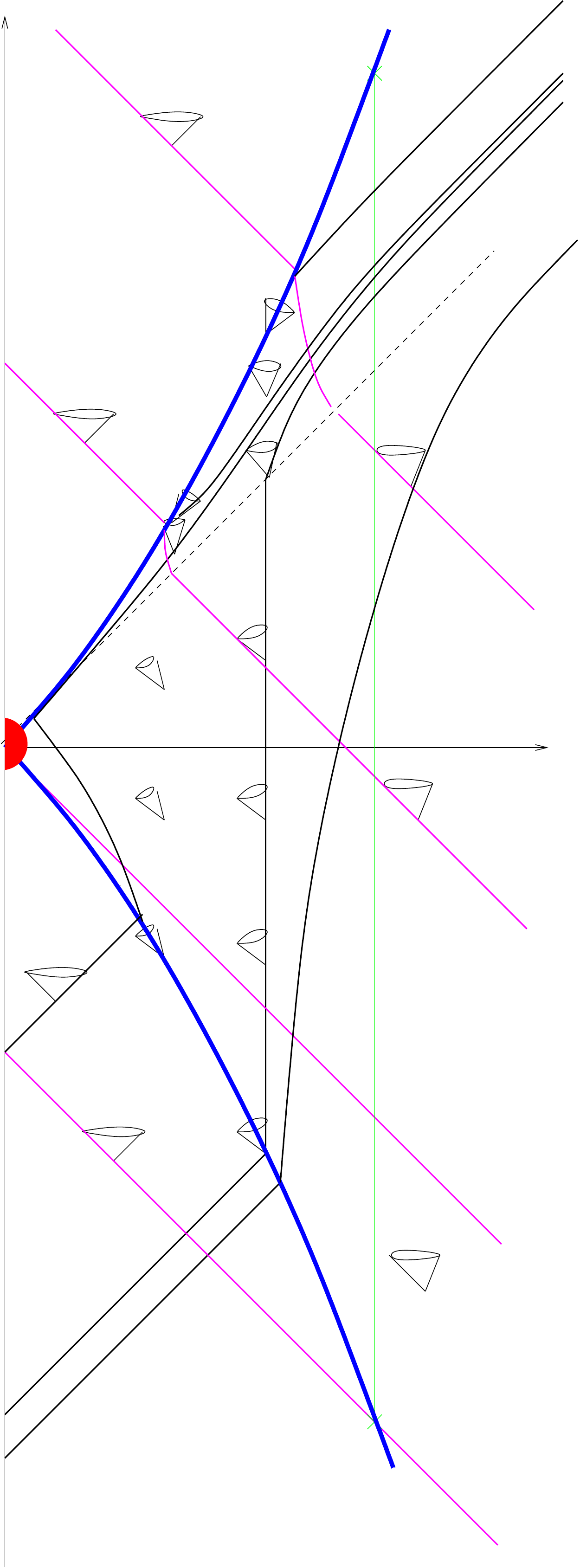}
\hfil}%
\bigskip%
\caption{\label{Fig:f-diagram}
Generalised Finkelstein diagram for a single collapse-and-bounce of a shell of matter (thick blue line) coming from spatial infinity.
The light-cone causal structure is displayed, together with some relevant
ingoing
(purple) and outgoing (black) null rays. In the central red patch, general
relativistic notions of curvature do not apply.
The vertical green line depicts the world-line of an observer at a fixed
radius.}
\end{figure}

The resulting geometry is shown schematically in Fig.~\ref{Fig:f-diagram}
in an extrapolation of the standard ingoing Finkelstein coordinates. 
The thick blue line represents the shell as it collapses and then bounces back.
For simplicity we draw a configuration with one single bounce, 
corresponding to a shell collapsing from spatial infinity. (For stellar configurations 
starting the collapse from a finite radius, there would be an infinite set of 
collapse-bounce-expansion cycles in the idealised case of absence of any dissipative and relaxation 
mechanism).
The small central red  patch represents the region in which gravity is not
operating and dynamics is governed exclusively by quantum field theory in
a flat spacetime. The thin black lines represent different outgoing null
rays. The dashed line represents the first outgoing causal signal
connecting the shell, just after its bounce, with the external world. The
region below the dashed diagonal line is geometrically identical to the
standard general relativistic formation of a black hole from a collapsing
shell. However, the existence of a bounce makes the extension of this
geometry above the dashed line completely different from the standard
case: it acquires features from a white-hole spacetime.

A crucial aspect of this geometry is the following. Consider the two
events $E1$ and $E2$ marked by green crosses in
Fig.~\ref{Fig:f-diagram} and two particular timelike curves connecting
them, corresponding to:
\begin{itemize}
 \item an observer $O1$ associated with a
free-falling observer attached to the shell (the thick blue line),
\item  a second observer $O2$ at rest at a fixed radius $r$
(the vertical green line) well outside the shell's Schwarzschild radius
$R_{\rm S}$: $r\gg R_{\rm S}$.
\end{itemize}
The time intervals between both  events  as measured by these two
observers are of the same order of magnitude. Indeed, the delay in the
collapsing phase observed by the outer observer at a fixed radius is
roughly compensated for by the rapid accumulation of all the light rays
emitted after the shell has crossed its Schwarzschild radius again on the
way out. The main essential difference between both time measurements
is a factor due to the gravitational field suffered by the outside
observer $O2$, which is very close to one for sufficiently distant
positions.

To illustrate this assertion, let us make an order-of-magnitude estimate of the time needed for a typical neutron star of two solar masses to collapse almost in free-fall
from an initial radius twice its Schwarzschild radius, and bounce
back. Take such a neutron star with $R_{\rm ns}=12$km and surface
gravity $a_{\rm ns}= g \times(R_{\rm earth}/R_{\rm ns})^2
 \sim 2.5 \times 10^6 {\rm m}/{\rm s}^2$. A Newtonian estimate
for the proper time of the free-fall collapse measured by observer $O1$ gives
\begin{eqnarray}
\tau_{\rm FF}=\sqrt{2 R_{\rm ns} \over a_{\rm ns}}\sim 0.1\text{s}~.
\end{eqnarray}
The symmetry of the process indicates that the total time for the
collapse and bounce (the time between the events $E1$ and $E2$ in
Fig.~\ref{Fig:f-diagram}) measured by the first observer is
$\tau_{O1}=2 \tau_{\rm FF}$. In terms of the Finkelstein time
\begin{eqnarray}
t_{\rm F}=t+2M\ln(r-2M)/2M~
\end{eqnarray}
(with $t$ the Schwarzschild time), if the `apparent singularity' is
located at $t_{\rm F}=0$, then the first event $E1$ (the onset of the
collapse) is located at \mbox{$t_{\rm F}=-t_{{\rm F}E1} \sim -\tau_{\rm
FF}$}. Again, for symmetry reasons, the total proper time between the
two events as measured by the second observer is
\mbox{$\tau_{O2} \sim (1-2M/r)^{1/2} \times (2t_{{\rm F}E1})$}. So, for
sufficiently large $r$, indeed, $\tau_{O1} \sim \tau_{O2}$.

\begin{figure}%
\vbox{\hfil
\includegraphics[width=.23\textwidth]{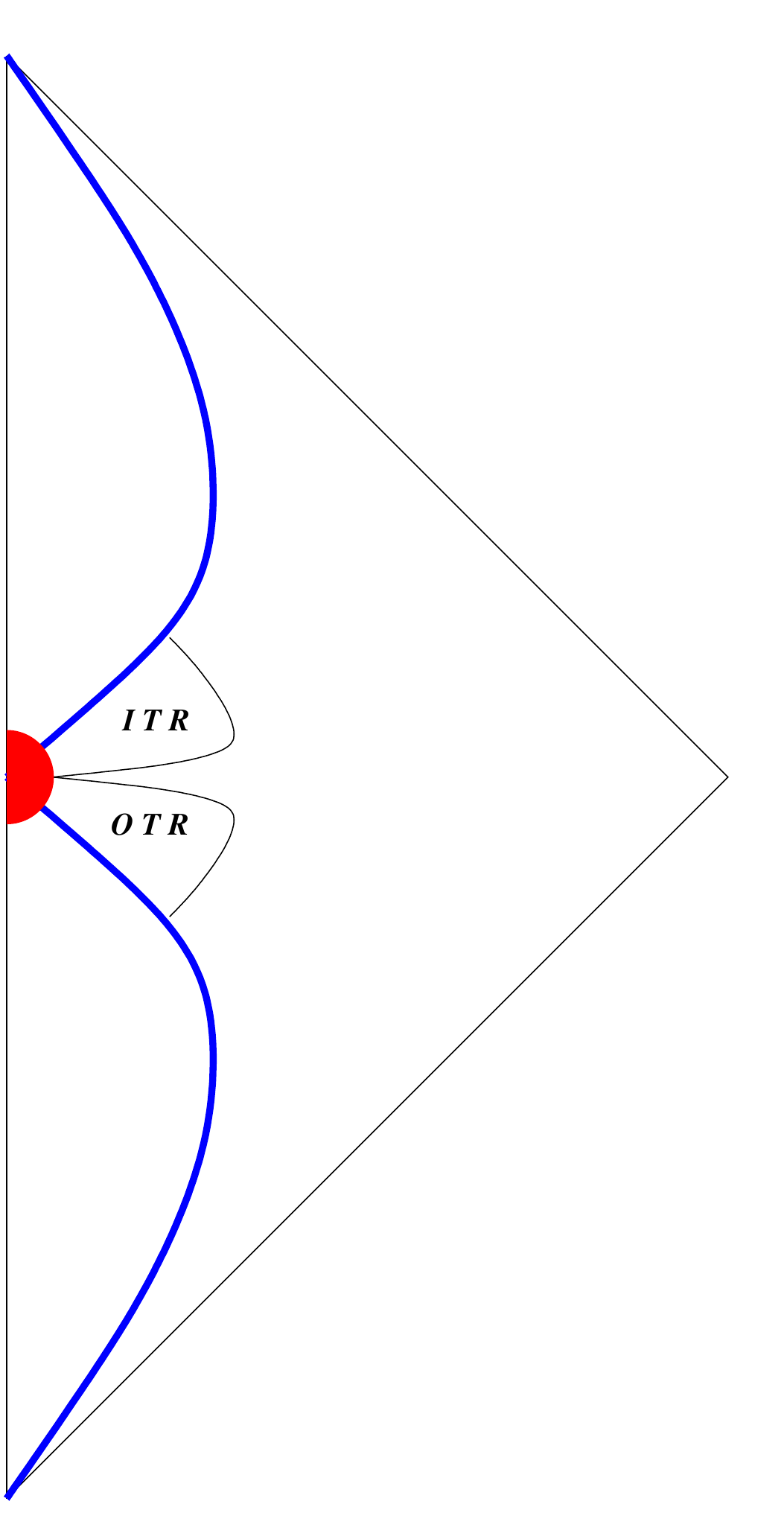}
\hfil}%
\bigskip%
\caption{Conformal diagram for our scenario (simple case of a single bounce). Causally, this spacetime is
equivalent
to Minkowski spacetime. General relativity breaks down in the central red
region.
The regions containing closed Future Trapped Surfaces (FTR) and closed
Past Trapped Surfaces (PTR) are also displayed.}
\label{Fig:c-diagram}%
\end{figure}

In Fig.~\ref{Fig:c-diagram} we have drawn the conformal diagram associated with the previous single-bounce geometry. Note that although this conformal diagram is similar to the one proposed by Ashtekar and Bojowald in~\cite{ashtekar-bojowald} (their Fig.~2), in the sense that they both contain a region where the classical notions of general relativity break down, they represent entirely different scenarios.
One difference can readily be seen in the causal structure depicted by the conformal diagrams.
The Ashtekar-Bojowald diagram contains a region of {\em future-trapped closed surfaces}.
In our simple one-bounce model there is in addition a region of {\em past-trapped closed surfaces}, which is absent in the Ashtekar-Bojowald diagram.  Considering now the entire geometry, not just the causal structure, it is not difficult to see that the `space-like' shaped
strong-gravity region in~\cite{ashtekar-bojowald} is intersected by a large set
of rays incoming from $\scri^-$. This is not the case for the nearly
`point-like' region in our geometry (see Fig.~\ref{Fig:f-diagram}). 
Remember that a single collapse-and-bounce process in our scenario lasts less 
than a second, as estimated above. In the Ashtekar-Bojowald proposal, 
on the other hand, the trapped regions disappear through an extremely slow 
(quasi-eternal) evaporation process. If one uses similar conformal compactification factors to draw both conformal diagrams, the relative size and shape of both strong gravity regions can also be seen by comparing both conformal diagrams. 

To summarize, contrarily to the diagram of
Ref.~\cite{ashtekar-bojowald}, our diagram does not represent a slowly evaporating black hole with a regular final phase. Rather, it represents a perfect bounce, and so just a
simplified description of a very brief transient stage in the evolution of the
collapsing matter towards a final non-black-hole equilibrium state. 
In particular, the formation of trapped surfaces displayed in
Fig.~\ref{Fig:c-diagram} will be part of the transient collapse-bounce epoch but
will not take part in the description of the final equilibrium state. 

\subsection{Multiple bounces and relaxation}

In this scenario, the whole stellar collapse process does not lead to a final stage consisting
of a slowly evaporating black hole. Instead, after several of the rapid bounces discussed above in a simple model,  the collapsing matter could reach an equilibrium, provided
that some dissipation and relaxation mechanisms are present (as will
always happen in any realistic model) such as
emission of gravitational radiation due to departures from sphericity and dynamical friction. 

We have to clearly distinguish between the collapse process and subsequent relaxation phase on the one hand, and the final resulting
(quasi-)stationary body on the other. Even if several of the previously
described rapid collapse-bounce phases were necessary to reach a
(quasi-)stationary configuration, the collapse-relaxation phase will last only a
few seconds as seen by distant observers. After that phase, the most reasonable result is to end up 
with a quasi-stationary astronomical body much like a compact and dark star,
with its surface very close but outside of the gravitational radius associated
with its entire mass and therefore, with no horizons of any kind. One could
consider the conformal diagram associated with an ``eternal idealization'' of such an
object in which it has always been there in the past and  will be there forever in
the future. This conformal diagram, equal to the trivial conformal diagram associated with Minkowski spacetime, is completely unrelated to the conformal diagram
in Fig.~\ref{Fig:c-diagram} which represents an idealized model of a single collapse-bounce
process in which the shell starts and ends at spatial infinity.

\subsection{The resulting body}

One possibility for such an equilibrium
configuration that has been examined in detail recently~\cite{barcelo-fate}, and
which fits nicely with the scenario that we are speculating on here, is a `black
star': A material body with a real and in principle  astrophysically explorable
surface, and a non-empty interior, filled with matter at least one order of
magnitude denser than neutron stars. As a consequence of its large gravitational
red-shift, such an object would be extremely dim and nearly indistinguishable
from a black hole  in the strict sense of general relativity. It would be
supported by the most basic form of quantum pressure: the one provided by the
quantum vacuum polarisation, which can be huge for configurations maintaining
themselves close to the formation of a horizon~\cite{barcelo-fate}. An
interesting aspect of this proposal is that it permits to `cure' general
relativistic singularities in a way that takes   the important notion of
isothermal spheres to its limit.  Indeed, the black stars just discussed can be
interpreted as limiting cases of isothermal spheres in the sense that the entire
structure maintains itself with a density profile
\begin{eqnarray}
\rho \lesssim {1\over 8\pi}{c^2 \over G r^2}~,
\end{eqnarray}
from its centre up to its surface which is located at a finite radius
$R_{\rm bs}$. This density profile represents a body in which the mass
within any radius $r<R_{\rm bs}$ is forever on the verge of forming a
horizon. Note that the central $1/r^2$ divergence is not there in the real
density profile but only in the limiting situation, which will never be
reached. Let us also mention that in~\cite{Barcelo:2006uw} it was shown
that black stars could emit Hawking-like radiation mimicking in this
respect evaporating black holes. Hawking radiative effects would make
each inner sphere to emit with a temperature approximately proportional
to the inverse of the mass enclosed in its radius $r$,
\begin{eqnarray}
T(r)\sim\frac{\hbar c^3}{8\pi G  M(r)}\sim \frac{\hbar c}{4\pi r}~,
\end{eqnarray}
and so increasing towards the centre. This would cause the inner volumes
to evaporate rapidly, further regularizing the real central density.

Although evaporating black holes with a regular final phase might not have a
strict event horizon, in astrophysical terms they can be described as hollows in
spacetime which last almost eternally. As we have discussed, the astrophysical
description of black stars is completely different. In the literature there are
other proposals of black hole mimickers similar in spirit to our proposal. Due to its closeness 
let us mention here Mottola-Mazur's gravastars~\cite{gravastars}. These objects don't have horizons either and they differ from our proposal in that their interior is a vacuum solution of the Einstein equations.

\subsection{A potentially observable consequence}

We end this section by presenting a potentially testable experiment
that would clearly distinguish black holes from black stars (or gravastars). The main feature that this experiment uses is that black stars have a real physical surface while the boundary of a black hole
is an event horizon. If we sent a radar signal straight towards a black hole it
would be completely absorbed (in the geometric approximation, at least) and hence no echo would return. In contrast, if the
same experiment were performed towards a black star whose surface were located at
a radius $r_s\gtrsim 2M$, then the time needed for the signal to go from an
observation point at $r_0$ to the black star surface and back would be given by
\begin{equation}
T=2\int_{r_s}^{r_0}\frac{\diff r}{1-2M/r}=2\left(r_0-r_s+2M\ln\frac{r_0-2M}{r_s-2M}\right).
\end{equation}
The relevant remark is that the general relativistic delay is logarithmic so
that, even though it diverges for a proper black hole, it  decreases to small
values very rapidly as the bouncing point departs from $2M$. For instance, for a
solar mass black star with a radius larger than its Schwarzschild radius ($3\times
10^3$m) by the tiny amount of $10^{-75}\mbox{m}$ (which is about $10^{-40}$
times the Planck length), a radar signal sent from a distance of 8 light-minutes ($=r_0/c$)
would acquire a gravitational delay of only $4$ milliseconds and would echo back
after about 16 minutes (plus 4 milliseconds), in sharp contrast with the infinite amount of time necessary in the case of a proper black hole.

\section{Conclusion} 

To summarise, we suggest that the first `quantum' correction
to gravity could be that gravity just switches off at high energies, leaving
essentially a Minkowskian quantum field theory. Of course, many questions with
respect to such a scenario remain to be explored. More detailed calculations
will be presented in future work. Here, however, we have already wished to
emphasize two points. First, such a suggestion is not as exotic as may seem at
first sight. Indeed, the underlying arguments are based on a combination of
observational indications and analyses from gravitational analogies in condensed
matter systems, more specifically from the well-established physics of Fermi
liquids. Second, these general considerations are sufficient to hint at the
possibility that evaporating black holes might not be the end-point of stellar
collapse. In the concrete example that we have discussed here, an equilibrium
situation could be reached in which the final object is a limiting case of an
isothermal sphere. We have shown that, although these object can mimic 
black holes in many respects, they have specific characteristics which would 
completely distinguish them from black holes.

\subsection*{Acknowledgements}

Financial support was provided by the Spanish MICINN through the
projects FIS2008-06078-C03-01 and FIS2008-06078-C03-03 and by
Junta de Andaluc\'{\i}a through the projects FQM2288 and FQM219. The
authors want to thank J.L. Jaramillo, S. Liberati, S. Sonego and M. Visser
for some illuminating discussions.




\begin{thebibliography}{99}
\bibitem{Jacobson:2005bg}
  T.~Jacobson, S.~Liberati and D.~Mattingly,
  ``Lorentz violation at high energy: concepts, phenomena and astrophysical
  constraints,''
  Annals Phys.\  {\bf 321} (2006) 150
  [arXiv:astro-ph/0505267].
\bibitem{Maccione:2009ju}
  L.~Maccione, A.~M.~Taylor, D.~M.~Mattingly and S.~Liberati,
  ``Planck-scale Lorentz violation constrained by Ultra-High-Energy Cosmic
  Rays,''
 JCAP {\bf 0904}, 022 (2009)
  [arXiv:0902.1756 [astro-ph.HE]].
\bibitem{Volovik:2006ft}
  G.~Volovik,
  ``From quantum hydrodynamics to quantum gravity,''
  in: H.~Kleinert, R.~T.~Jantzen and R.~Ruffini (eds.), {\it Proceedings of the
11th Marcel Grossmann Meeting on General Relativity,} World Scientific,
Singapore (2008) [arXiv:gr-qc/0612134].
\bibitem{Volovik:2003fe}
  G.~E.~Volovik,
  {\it The Universe in a helium droplet,}
  Clarendon Press, Oxford (2003).
\bibitem{Volovik:2007vs}
  G.~E.~Volovik,
  ``Fermi-point scenario for emergent gravity,''
  PoS QG-Ph:043 (2007) [arXiv:0709.1258 [gr-qc]].

\bibitem{atiyah} M.F. Atiyah, R. Bott and A. Shapiro, ``Clifford Modules,'' Topology 3 Suppl. {\bf 1}, 3 (1964).
\bibitem{Horava:2005jt}
  P.~Ho\v{r}ava,
  ``Stability of Fermi surfaces and K-theory,''
  Phys.\ Rev.\ Lett.\  {\bf 95}, 016405 (2005).
  [hep-th/0503006].


\bibitem{Sakharov:1967pk}
  A.~D.~Sakharov,
  ``Vacuum quantum fluctuations in curved space and the theory of gravitation,''
  Sov.\ Phys.\ Dokl.\  {\bf 12}, 1040 (1968)
  [Dokl.\ Akad.\ Nauk Ser.\ Fiz.\  {\bf 177}, 70 (1967)].
\bibitem{Visser:2002ew}
  M.~Visser,
  ``Sakharov's induced gravity: A modern perspective,''
  Mod.\ Phys.\ Lett.\  A {\bf 17} (2002) 977
  [arXiv:gr-qc/0204062].
\bibitem{Barcelo:2005fc}
  C.~Barcel\'{o}, S.~Liberati and M.~Visser,
  ``Analogue gravity,''
  Living Rev.\ Rel.\  {\bf 8}, 12 (2005)
  [arXiv:gr-qc/0505065].
http://www.livingreviews.org/lrr-2005-12.
\bibitem{Weinberg:1980kq}
  S.~Weinberg and E.~Witten,
  ``Limits On Massless Particles,''
  Phys.\ Lett.\  B {\bf 96} (1980) 59.
\bibitem{Boughn:2008jx}
  S.~Boughn,
  ``Nonquantum Gravity,''
 Found.\ Phys.\  {\bf 39}, 331 (2009)
  [arXiv:0809.4218 [gr-qc]].

\bibitem{barcelo-fate}
  C.~Barcel\'{o}, S.~Liberati, S.~Sonego and M.~Visser,
  ``Fate of gravitational collapse in semiclassical gravity,''
  Phys.\ Rev.\  D {\bf 77}, 044032 (2008)
  [arXiv:0712.1130 [gr-qc]].
\bibitem{Barcelo:2006uw}
  C.~Barcel\'{o}, S.~Liberati, S.~Sonego and M.~Visser,
  ``Hawking-like radiation does not require a trapped region,''
  Phys.\ Rev.\ Lett.\  {\bf 97}, 171301 (2006)
  [arXiv:gr-qc/0607008].
\bibitem{ashtekar-bojowald}
  A.~Ashtekar and M.~Bojowald,
  ``Black hole evaporation: A paradigm,''
  Class.\ Quant.\ Grav.\  {\bf 22}, 3349 (2005)
  [arXiv:gr-qc/0504029].
\bibitem{gravastars}
  P.~O.~Mazur, E.~Mottola,
  ``Gravitational vacuum condensate stars,''
  Proc.\ Nat.\ Acad.\ Sci.\  {\bf 101}, 9545-9550 (2004).
  [arXiv:gr-qc/0407075].




\end{thebibliography}
\end{document}